# Real-time Over-the-air Adversarial Perturbations for Digital Communications Deep Neural Networks

Roman A. Sandler, Peter K. Relich, Cloud Cho, Sean Holloway

*Abstract*—**Deep neural networks (DNNs) are increasingly being used in a variety of traditional radiofrequency (RF) problems. Previous work has shown that while DNN classifiers are typically more accurate than traditional signal processing algorithms, they are vulnerable to intentionally crafted adversarial perturbations which can deceive the DNN classifiers and significantly reduce their accuracy. Such intentional adversarial perturbations can be used by RF communications systems to avoid reactive-jammers and interception systems which rely on DNN classifiers to identify their target modulation scheme. While previous research on RF adversarial perturbations has established the theoretical feasibility of such attacks using simulation studies, critical questions concerning real-world implementation and viability remain unanswered. This work attempts to bridge this gap by defining class-specific and sample-independent adversarial perturbations which are shown to be effective yet computationally feasible in real-time and time-invariant. We demonstrate the effectiveness of these attacks across a physical channel using software-defined radios (SDRs). Finally, we demonstrate that these adversarial perturbations can be emitted from a source other than the communications device, making these attacks practical for devices that cannot manipulate their transmitted signals at the physical layer.**

## I. INTRODUCTION

IN the last several years deep neural networks (DNNs) have revolutionized many problems in computer vision and natural language processing. Increasingly, DNNs are being applied to several problems in RF communications including RF fingerprinting [1], spectrum sensing [2], demodulation [3], and automatic modulation classification (AMC) [4]. As in computer vision, these studies have consistently shown that DNNs have superior performance than traditional signal processing algorithms.

However, as originally shown in computer vision, DNNs are vulnerable to so-called adversarial perturbations which subtly perturb the input sample so as to cause the classifier to misclassify [5]. In computer vision this perturbation is constrained to be imperceptible to humans. In RF communications, which are not typically subject to human analysis, the perturbations are constrained to not interfere with signal decoding.

Several previous studies have shown that, like their computer vision correlates, DNNs used in RF are also vulnerable to adversarial attacks [6-7]. This opens intriguing attack vectors on communications networks that make use of these DNN classifiers. Most previous RF adversarial attack studies have focused on automatic modulation classification (AMC) algorithms due to the availability of suitable open-source datasets. Such AMC algorithms are typically found in the early stages of reactive jammers and interception systems and 'activate' follow-on action (e.g., jamming or interception) once the desired modulation scheme is detected. As DNNs are increasingly considered for use as AMC algorithms due to their superior accuracy, these systems will be vulnerable to adversarial attacks. Thus, an adversary may add perturbations to their communications signals to deceive the opposing AMC classifier and thus prevent it from triggering interception and/or jamming.

While multiple previous studies have established the theoretical feasibility of adversarial attacks in RF, there are multiple issues regarding the practical implementation and effectiveness of such attacks. Specifically, the generation of most adversarial perturbations is computationally intensive and cannot be done at latencies sufficiently low for RF communications. Furthermore, while these attacks have been shown under ideal and simulated channel conditions, to our knowledge, they have not yet been demonstrated over-the-air (OTA), where real-life complex channel conditions may potentially 'wash-away' and render ineffective the subtle adversarial perturbations. This paper aims to address these issues by:

1) Defining class-specific universal adversarial attacks (CUAAs) which are more effective than fully universal adversarial attacks while still being able to be performed in real-time because they are not optimized for specific samples but for the sample classes.

2) Analyzing universal, class-specific, and sample-specific adversarial targeted attacks.

3) Showing that adversarial attacks remain effective when broadcast OTA across a real channel. We generate a custom dataset where the ground truth bit message is known. We use this dataset to show that the adversarial attack can not only deceive a DNN classifier OTA but may do so while still being decodable with low bit-error rate (BER). We demonstrate this over multiple distances in both line-of-sight (LoS) and non-Line-of-Sight (NLoS) channels. To our knowledge, we are the first to demonstrate this.

Additionally, this paper defines 'smart jamming' whereby universal adversarial perturbations are broadcast on a separate device from the signal-transmitting device (Figure 1). This capability allows adversarial attacks to be carried out alongside legacy RF devices which have no means to manipulate their transmitted signals on the physical layer. It enables a single attack device to 'cloak' multiple allied transmit devices and deceive multiple opposing RF classifiers in the vicinity. These classifiers are rendered ineffective at a much lower transmit

Authors are with Intellisense Systems, Inc, Torrance, CA. E-mail: (rsandler,prelich,ccho,sholloway)@intellisenseinc.com. This work was funded by Navy contract N6833520G1021 power than would be needed by a traditional jammer. Unlike standard jammers, smart jamming does not impede desired communication at its carrier frequency. Such a capability would be useful, for example, for a drone swarm where most drones may not have the capability to inject adversarial perturbations into their physical layer in-phase and quadrature (IQ) transmissions. Nonetheless, a single drone in the swarm may contain a more sophisticated payload such as an SDR and broadcast these perturbations on behalf of the entire swarm and thus impede any RF DNN-based classifiers used to identify the drones for follow-on eavesdropping or jamming.

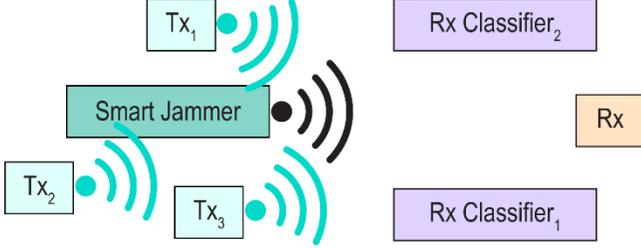

*Figure 1. Smart jamming concept of operations (CONOPS). Multiple transmitter devices (Tx$_n$) are communicating with a receiver (Rx). A smart jammer is located among the transmitting devices and broadcasts universal adversarial perturbations. The adversarial perturbations do not interfere with signal decoding on the Rx device but deceive the classifier on the Rx classifier devices.*

## II. PRIOR WORK

Deep learning (DL) has revolutionized many problems in computer vision [8]. O'shea et al., 2017 [3-4] was one of the first works to apply DL to the physical layer for the purposes of modulation classification. The ResNet architecture and RadioML dataset used in this work has subsequently been adopted as a baseline architecture and dataset by many works in the field including this one.

Szegedy et. al., 2013, demonstrated the first use of adversarial attacks. Since then, many novel attacks and defenses have been proposed, as reviewed in Ref. [9]. Today, while there exist many defenses for specific attacks, there is no defense with theoretical guarantees of security, and defenses that succeed against specific attacks are often shown to succumb later to new or modified attacks [10].

Sadeghi and Larsson [6] were the first to demonstrate adversarial attacks against RF (AMC) classifiers. Their work explored the use of universal, sample-independent, adversarial attacks. Flowers et al., 2019, was the first study to verify that these adversarial perturbations do not impede signal decoding as measured by BER. Bair et al., 2019, examined targeted adversarial attacks in RF, whereby the adversarial perturbation is optimized to make the classifier misclassify the input sample as a specific class. Hameed et al., 2020, studied the efficacy of such adversarial perturbations in more realistic conditions by simulating transmission over a Gaussian channel. Berian et al., 2020, showed that universal RF adversarial perturbations can be generated via filtering as opposed to addition.

Much less work has been done on defending RF classifiers against adversarial attacks. Recently, Sahay et al., 2021, have shown that an ensemble of time-domain and frequency domain classifiers are robust against IQ-generated adversarial attacks. Kokalj-Filipovic et al., have proposed to defend against adversarial attacks using autoencoder pretraining. Kim et al., 2021, have proposed a defense based on randomized smoothing [11]. As in computer vision, none of these defenses have any theoretical guarantees of robustness, and it is unknown how they will stand up to future attacks.

## III. METHODS

### A. DNN Data, Architecture, and Training

Here we define the mathematical notation used to distinguish the different types of adversarial attacks we analyze in this work. Let $\{\mathbf{x}_n, \mathbf{y}_n\}$ denote a dataset of N labeled examples. $\mathbf{x} \subset \mathbf{R}^p$ is the input data with $p$ being the dimensions of the inputs, and $\mathbf{y} \subset \mathbf{R}^c$ with $C$ being the number of output classes. In our case, $\mathbf{X}_n$ corresponds to raw IQ samples and $\mathbf{y}_n$ corresponds to the modulation scheme of those samples. Let $f(., \boldsymbol{\theta}) : \mathbf{x} \rightarrow \mathbf{R}^C$ denote a DNN classifier parametrized by weights $\boldsymbol{\theta}$. For every input $\mathbf{x}$ the classifier assigns a label $l(\mathbf{x}, \boldsymbol{\theta}) = \arg \max_k f_k(\mathbf{x}, \boldsymbol{\theta})$ where $f_k(\mathbf{x}, \boldsymbol{\theta})$ is the output of $f$ corresponding to the $k$th class.

In our context of automatic modulation classification, for $f$, the ResNet architecture was adopted from Ref. [4] with some modifications that were found to improve results. Namely, Batch Normalization is used only before the max pooling layer of each residual stack and after each of dense layers excluding the last. Standard dropout (dropout rate = .5) is used prior to each of the dense layers excluding the last. To increase data diversity and reduce overfitting, during training random Gaussian noise (from 0 to 30 dB signal-to-noise ratio, SNR) and random phase offsets (from -180° to 180°) are added to each batch. The latter was found to be particularly crucial for successfully classifying OTA data.

Two datasets were used in this work. First, following much of the work in this area, the RadioML dataset [4] was used for training. RadioML's input data consists of 1024 sample raw IQ data ($p = [1024, 2]$) and $C = 24$ output modulation scheme classes. The RadioML dataset does not include the underlying encoded bits encoded by the signal. Thus, it is impossible to analyze to what extent adversarial perturbations impede signal decoding. To study this, a custom modulation classification dataset was generated using Matlab's Communication Systems Toolbox. Four modulation schemes were used: PSK, APSK, QAM, and FSK, each at multiple common modulation orders for a total of 16 modulation schemes: bpsk, qpsk, 8psk, 16psk, 16apsk, 32apsk, 64apsk, 16qam, 32qam, 64qam, 128qam, bfsk, qfsk, 8fsk, 16fsk. We note that higher-order modulation schemes are inherently more difficult to decode and are expected to have higher BER. For each scheme, a rate of 4 samples-per-symbol was used and a pulse-shaping filter was applied using a root-raised-cosine shape. Python 3.8 and Tensorflow 2 was used for all development.

### A. Adversarial Attacks

Three types of adversarial attacks were used in this work: probabilistic gradient descent (PGD) [12], universal adversarial attacks (UAA) [13], and class-specific universal adversarial attacks (CUAA) [14]. Standard adversarial examples seek to fool a given DNN classifier by constructing an adversarial perturbation $\mathbf{r}_x$ for a specific input sample $\mathbf{x}$ such that (1) the classifier misclassifies the perturbed input $(\mathbf{x} + \mathbf{r}_x)$ (2) the perturbation is sufficiently 'small':

$$\min \|\mathbf{r}_x\|_p \text{ s.t. } l(\mathbf{x}, \boldsymbol{\theta}) \mathrel{!=} l(\mathbf{x} + \mathbf{r}_x, \boldsymbol{\theta}) \qquad (1)$$

Many methods have been proposed to solve the above optimization problem. Here, we use the PGD method, which iteratively uses backpropagation to move the input, **x**, in a direction that maximizes the classifier loss function while constraining the perturbed input to be within a ball with preset radius, $\varepsilon$ [12].

Standard adversarial attacks such as PGD that minimize Eq. 1 are computationally expensive since multiple rounds of backpropagation must be applied to each input. This is impractical for RF applications due to the very high throughput rates of modern RF devices (e.g., 10 MS/sec). To overcome this obstacle, we follow [sag] and adopt sample-independent UAAs that aim to identify an adversarial perturbation that fools the classifier for arbitrary inputs:

$$\min \|\mathbf{r}\|_p \text{ s.t. } l(\mathbf{x}, \boldsymbol{\theta}) \mathrel{!}= l(\mathbf{x} + \mathbf{r}, \boldsymbol{\theta}) \tag{2}$$

Note that once the adversarial perturbation, **r**, is found, the only 'real-time' operation to be performed is addition, which makes this method amenable to practical applications. We use a modified version of the original algorithm described in Ref. [13] to craft UAAs. Briefly, the method iteratively generates adversarial perturbations from random inputs selected from the training set and adds them to each other while constraining them to a ball with preset radius, $\varepsilon$. Our main modification was to add random phase offset to the sampled RF data prior to each iteration. We found this was crucial for making the adversarial perturbations robust over a real physical channel in OTA transmissions.

An additional benefit of the UAA approach is that since the adversarial perturbation is optimized for 'arbitrary' inputs, it is time-invariant (i.e., unsynchronized) with respect to the input. This enables one to transmit the adversarial perturbation from a separate emission source than the actual signal without performing any synchronization protocols. We explore this capability, which we call 'smart jamming'.

We have also implemented what we call class-specific UAAs (CUAAs) whereby a different universal adversarial perturbation is generated for each input class. This hybrid approach between UAAs and standard adversarial attacks is specific with respect to input class but universal with respect to input sample. In practice, this adds little computational overhead since in RF communications, modulation schemes are expected to be changed relatively infrequently, and when they do change, the only additional operation would be loading the correct perturbation into the transmission pipeline.

In addition to standard 'untargeted' adversarial attacks where the objective is to cause the classifier to misclassify (e.g., $l(\mathbf{x}, \boldsymbol{\theta}) \mathrel{!}= l(\mathbf{x} + \mathbf{r_x}, \boldsymbol{\theta})$), we have also studied targeted adversarial attacks where the objective is to cause the classifier to misclassify input samples as specific output classes (e.g., $l(\mathbf{x}, \boldsymbol{\theta}) = t$), where $t$ is the target class. For PGD, this is done by setting the backpropagation loss to be the difference between the actual classifier outputs, $l(\mathbf{x}, \boldsymbol{\theta})$ and the target class represented as a one-hot vector.

We used the CleverHans library implementation of PGD and developed custom implementations of UAA and CUAA attacks based on PGD. Unless otherwise stated, in all subsequent analysis, the PGD ball was clipped based on the $l_2$ norm, the number of PGD iterations was set to 50, and the step size of each attack iteration was set to 0.05.

### B. OTA Analysis

To verify that the adversarial attacks we are analyzing function OTA, across real channel conditions, we used two SDRs: an Ettus N200 (Rx) and Ettus X300 (Tx). Both SDRs were positioned to be within LoS distance from each other at a distance of 4ft unless otherwise noted. All transmission were done at 5GHz center frequency with a 10 MHz bandwidth. To avoid carrier frequency offset, which is outside the scope of the current work, an SMA cable was used to synchronize both radios' clocks. GNURadio was used for Tx/Rx data collection.

To study signal decoding over a real channel, a basic receiver is simulated. The correlation is calculated between the transmitted signal and the received test. The location, amplitude, and phase of the maximum correlation value are used to recover the timing offset and phase offset of the signal under test. This re-normalized signal is then demodulated, and the BER is calculated. It should be noted that this technique cannot handle all channel effects as a state-of-the-art receiver would, but is useful for comparison between the BER of different cases.

### IV. RESULTS

#### A. DNN Training

The ResNet network was successfully trained on the RadioML dataset. Average accuracy in low-noise conditions (>10 dB SNR) was 82.5% and was 51.2% at 0 dB SNR (Figure 2). Both values are superior to the results in the original RadioML paper [4] showing the effectiveness of our network architecture modifications and training data augmentation procedures. The same network architecture and training procedure was also used to successfully train the internally generated dataset to >99% accuracy (Figure 9).

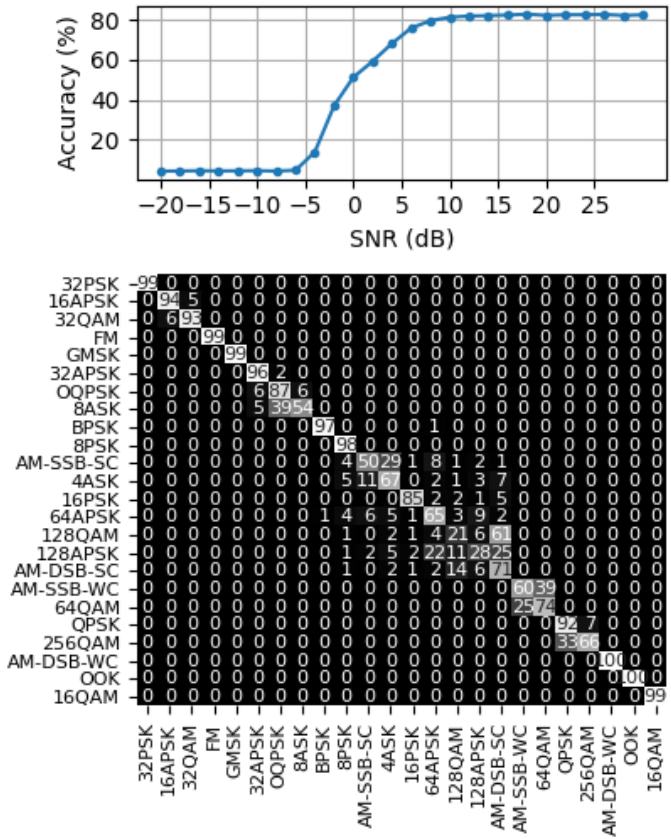

minimal effect. As expected, the PGD attack, which customizes perturbations for each specific input, was more effective than the UAA and CUAA attacks. The CUAA attack, which customizes adversarial perturbations for each input class, was the found to be more effective than the standard UAA attack and even than the PGD attack at low epsilon values.

*Figure 2. RadioML classifier results: (A) Classifier accuracy vs SNR. (B) Confusion matrix of all examples with SNR > 0 dB.*

### B. Adversarial attacks

After the network was trained, all three types of adversarial attacks were generated. Figure 3 shows the generated UAA adversarial perturbation using 512 examples, and an $l_2$ norm with $\varepsilon$ set to 8. Figure 4 shows two UAA examples where the same adversarial perturbation from Figure 3 was applied to two examples with ground truth modulations OOK (left) and 128 APSK (right). The perturbation caused them to be misclassified as 32 APSK and 4 ASK, respectively.

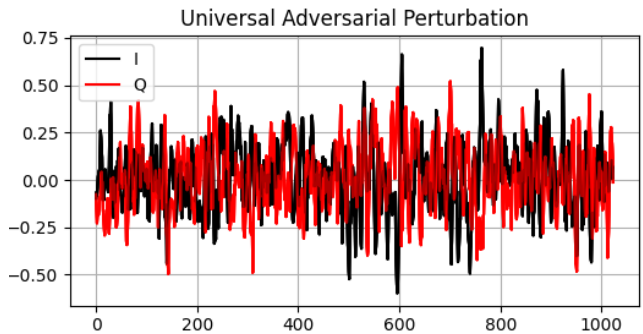

*Figure 3. Universal adversarial perturbation.*

To compare the effectiveness of all three attacks, we quantified their accuracy on 'clean' (>25 dB SNR) examples from the RadioML dataset over increasing values of $\varepsilon$. (Figure 5). As a control, we included both Gaussian white noise (GWN) of the same magnitude as the adversarial perturbations and a randomly shuffled UAA mask (UAA_Pert). All three adversarial attacks managed to substantially reduce classifier accuracy, while control noise (GWN and UAA_Pert) had

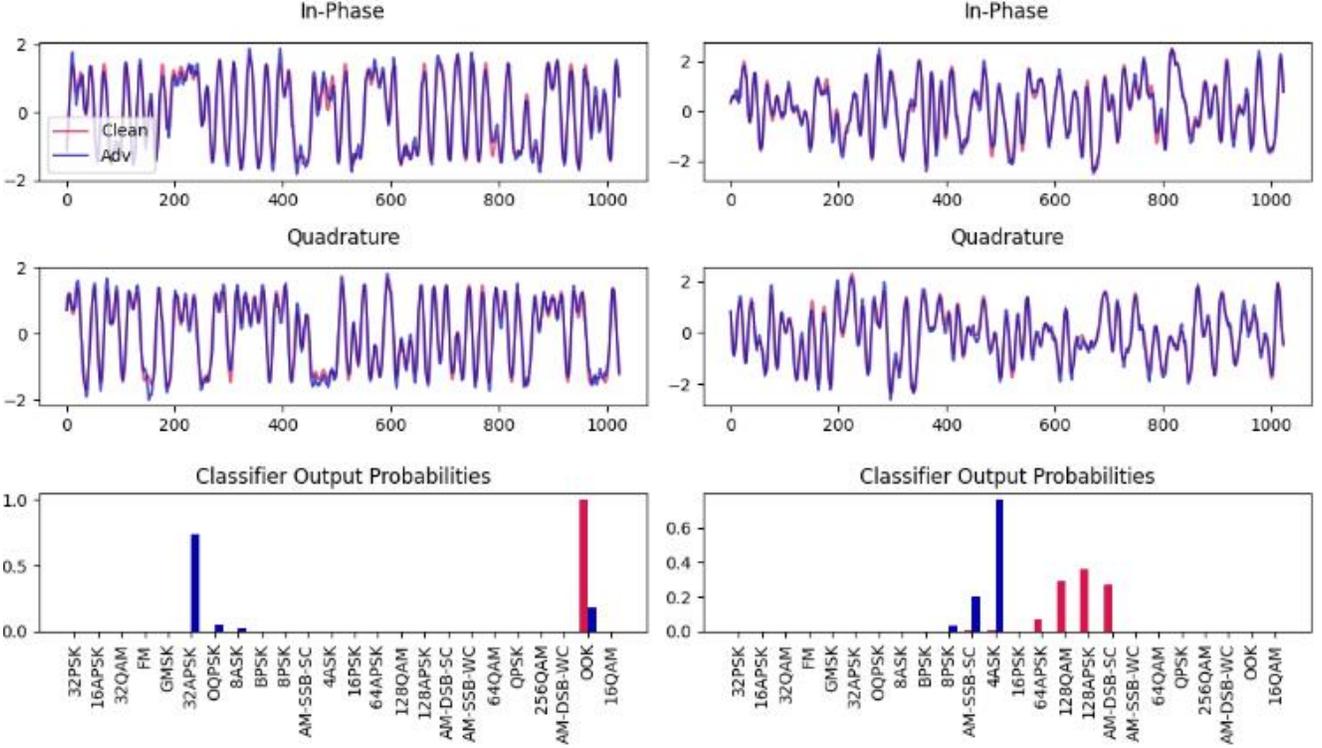

*Figure 4. Left and right columns show two separate UAA attack examples. (A) and (B) show the original ('clean') and adversarial data for in-phase and quadrature components. Adversarial attacks for both examples were generated with the adversarial mask from Figure 1. (C) shows the classifier outputs for both examples, with bar height representing classifier output probabilities, $l(\mathbf{x}, \boldsymbol{\theta})$. Note both adversarial examples are misclassified.*

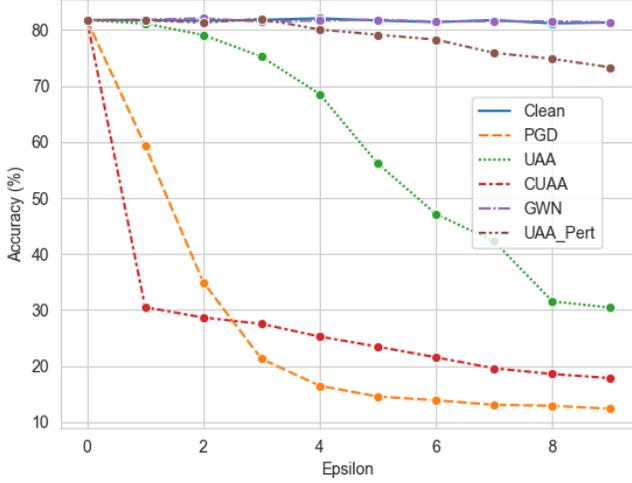

*Figure 5. Accuracies of various attacks and noise controls across increasing magnitudes.*

The above results analyze adversarial attack effectiveness without considering real-world channel propagation effects. To our knowledge, only the work of Hameed et al., 2020 [15], has considered such effects and we seek to expand those investigations here. In preparation for our OTA analysis, we analyzed adversarial attack efficacy when channel noise is added *after* the adversarial perturbation. We used a basic GWN model of channel effects, and quantified post-adversarial noise (PAN) as:

$$\sigma_p = 20log\frac{\sigma_a}{\sigma_{GWN}} \text{ dB} \tag{3}$$

where $\sigma_a$ is the variance of the adversarial perturbation. The results are shown in Figure 7. As can be seen, to a certain extent, adding Gaussian noise, and thus decreasing $\sigma_p$, decreases adversarial attack efficacy and improves classifier performance. This is because the adversarial perturbations are subtle by design so as to not interfere with signal decoding, and thus can potentially be 'washed away' by channel noise. Interestingly, both UAA and CUAA attacks are fully 'washed away' with increased noise and converge to identical performance as standard noise ('Clean' in Figure 6), while the PGD attack is still effective with large amounts of channel noise (0 dB $\sigma_p$).

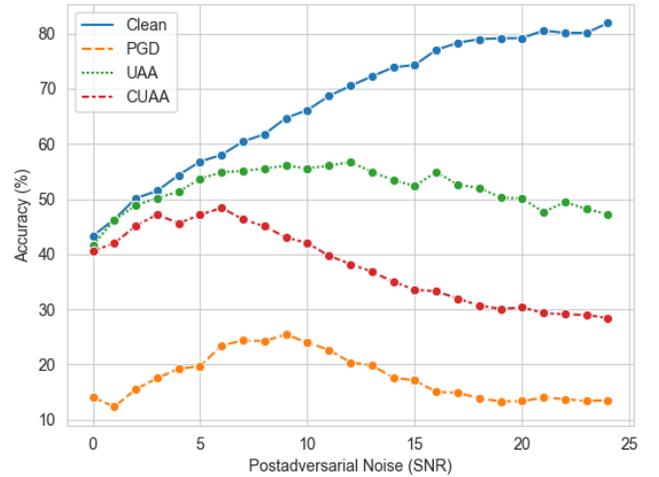

*Figure 6. Adversarial attack efficacy, as measured by classifier accuracy, when GWN of various magnitudes is added after the adversarial perturbation.*

## C. Targeted Adversarial attacks

To further understand the nature of UAA and CUAA attacks, we studied targeted adversarial attacks where the adversarial perturbations were not simply optimized to cause the classifier to misclassify, but rather to classify the input sample as a specific (incorrect) target class. In our analysis, we generated targeted adversarial attacks for each possible target class, $t$, of which there are 24 in the RadioML dataset. To quantify targeted adversarial attack efficacy, the following metric was used:

$$A(s,t) = \frac{\sum_{x_s} \mathbf{1}[l(x+r)=t]}{N_s} \qquad (4)$$

Where $t$ is the target class, $s$ is the source class, $N_s$ are the amount of source class samples, and $\mathbf{1}[.]$ is the indicator function. The targeted adversarial attack results for each of the studied adversarial attacks are shown in Figure 7. As can be seen, sample-independent attacks (UAA and CUAA) were far less effective than the PGD for targeted attacks. Both UAA and CUAA attacks pushed the classifier towards a small set of output classes (in this case, AM-DSB-SC and to a lesser extent 128 QAM). Even with the more successful PGD attack, some source classes (e.g., BPSK) and target classes (e.g., 16 PSK) were much more amenable to targeted attacks than others.

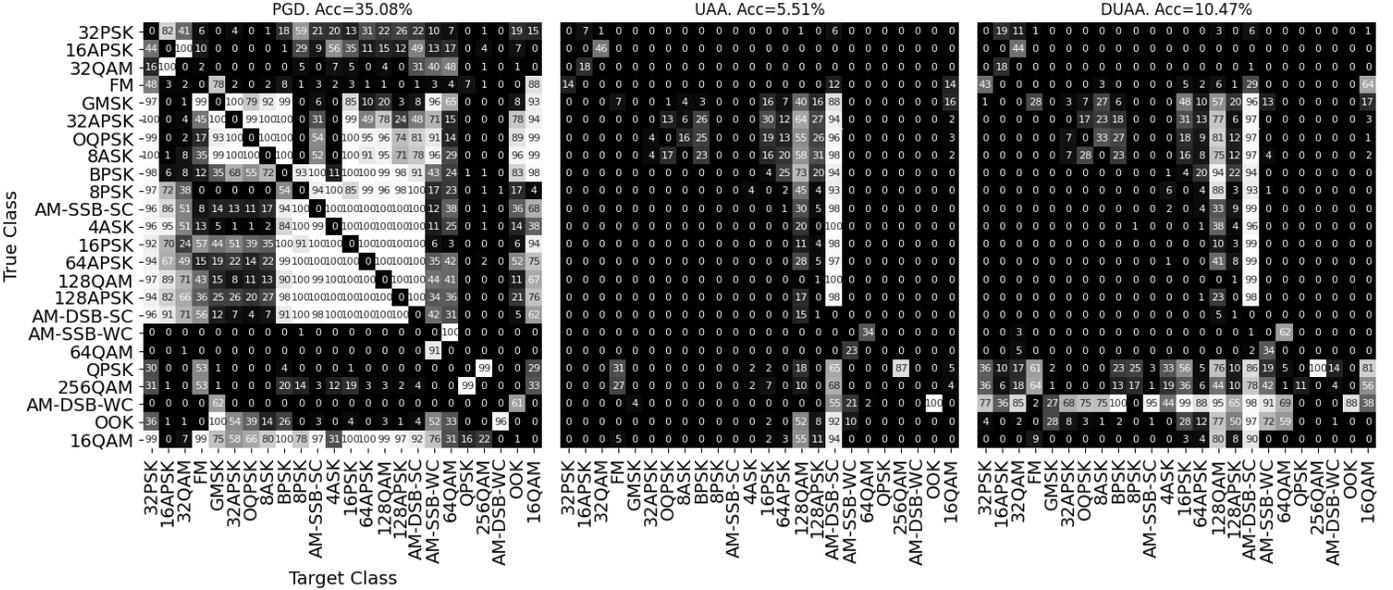

*Figure 7. Adversarial attack results for PGD, UAA, and CUAA attacks. Each value shows A(s, t) for the respective source and target classes. Diagonal entries were set to 0 to preserve scale.*

## D. OTA Adversarial Attacks

To verify that adversarial attacks work OTA and not just in simulation, SDRs were used to broadcast samples of both datasets with and without adversarial noise across two SDRs within LoS at varying distances from 4 to 16 ft. An additional test was conducted at 16 ft NLoS. The trained classifiers for each dataset were then used to classify both the simulated Tx signal and the OTA Rx signal. The results for the RadioML dataset across different SNRs are shown in Figure 8. Only high SNR (>25 dB) data from the RadioML dataset was used in OTA transmissions. Classifier performance dropped by 5.4% and 3.7% between the simulated and 4 ft OTA data for clean and adversarial data, respectively. We believe this is due to the classifier being trained exclusively on simulated data and not fine-tuned on OTA data. Nonetheless, despite the signal degrading by 11.8 dB over the range of the tests and the addition of multipath fading effects, classifier performance dropped by <1% for both clean and adversarial signals, showing that at least in the range tested, real channel effects do not reduce the efficacy of either the trained DNN classifier or of the adversarial examples.

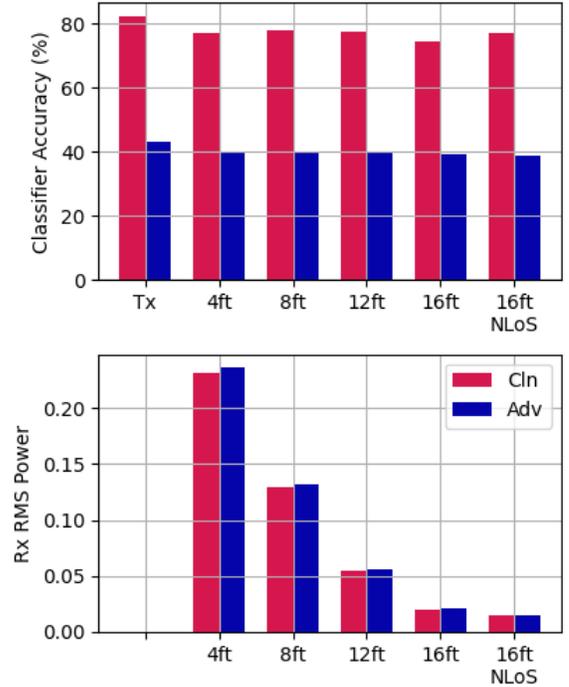

*Figure 8. (Top) Classifier accuracy on simulated (Tx) and OTA (Rx) clean and adversarial samples from the RadioML dataset across a range of distances. (Bottom) Received signal RMS power.*

The classifier results for the custom dataset are shown in Figure 9. Signals with artificially injected noise were also analyzed to ensure that the decline in classifier accuracy seen with adversarial perturbations were not simply the results of added noise. Specifically, Gaussian noise of the same magnitude as the perturbations (GWN) as well as randomly permuted versions of the perturbations (Per) were transmitted. Unlike the adversarial perturbations, neither of these manipulations substantially decreased accuracy.

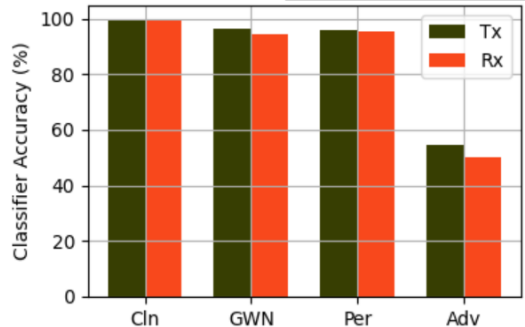

*Figure 9. Classifier accuracy on simulated (Tx) and OTA received (Rx) data. Data was transmitted with (Adv) and without (Cln) adversarial perturbations, as well as noise controls (GWN, Per).*

### E. OTA Adversarial Decoding

With the exception of Ref. [7], most work that examines RF classifier adversarial perturbations only looks at SNR but does not explicitly examine the effects of the perturbations on decoding. This is partly because the widely used RadioML dataset does not include the ground truth bit message of the signals (and in fact includes several analog modulation schemes). The internally generated modulation data set was used to specifically study the effects of adversarial perturbations on decoding and to compare these perturbations with other forms of noise.

Figure 10(a) shows decoding results for simulated modulated data with varying amounts of AWGN (dashed lines) and adversarial perturbations (dashed lines) added. As can be seen, adversarial perturbations induce slightly more decoding error than comparable AWGN, especially at high SNRs. This phenomenon is likely due to either beat frequencies caused by the repeating adversarial perturbation or some correlation between perturbations that decrease classifier performance and decoding efficacy.

Decoding efficacy of adversarial perturbed signals was also examined after the signals were broadcast OTA. Here, the adversarial perturbation was fixed to be 14 dB below the clean signal. The decoding BER across modulation schemes for clean, noisy, and adversarial perturbed signals is shown in Figure 10(b). As before, noisy signals include Gaussian noise of the same magnitude as the perturbations (GWN) as well as randomly permuted versions of the perturbations (Per). As expected, higher order modulation schemes have higher BER. Our intent here was not to minimize BER but to analyze how much more decoding error, as measured by BER, is caused by adversarial perturbations as opposed to noise of equivalent magnitude. Figure 10(c) shows BER averaged across modulation schemes. As can be seen, the adversarial perturbation increases BER by <1% compared to equivalent AWGN signals. This is consistent with the trend seen in simulated data (Figure 10(a)) and confirms the viability of using adversarial perturbations in real communications systems.

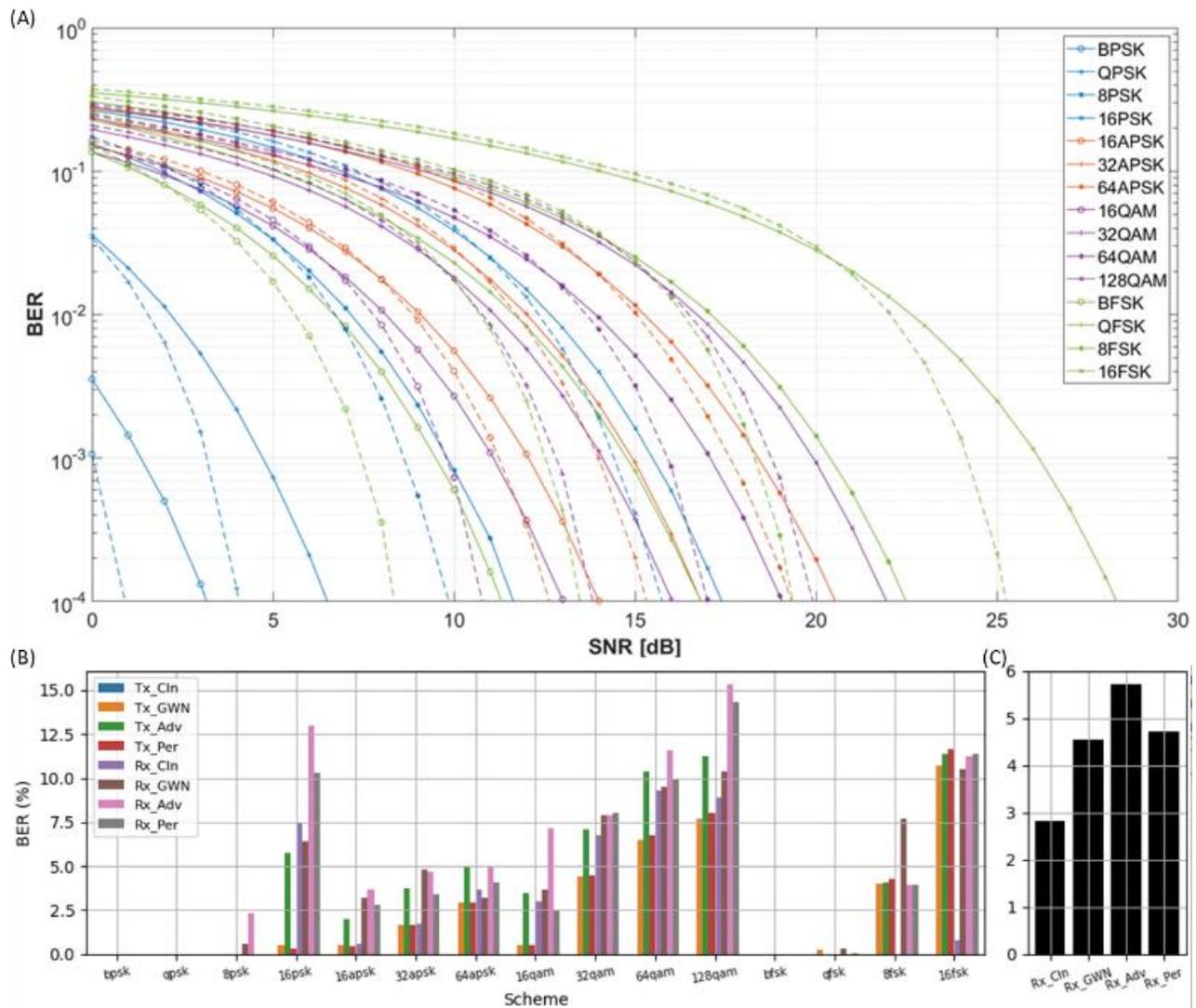

*Figure 10. Comparison of BERs for all examined modulation schemes on simulated data with variable amounts of AWGN (dashed lines) and adversarial perturbations (solid lines). (B) and (C) Show transmitted and OTA received BER (%) for each modulation scheme and averaged across all modulation schemes, respectively.*

### F. Smart Jamming

An advantage of sample-independent adversarial attacks is that they do not have to be synchronized with input data – they are time invariant. Thus, in Eq. (2), there can be an arbitrary time delay between **x** and **r**, and, theoretically, attack efficacy should not decrease. This introduces the capability we call 'smart jamming' where an independent transmitter can 'jam' target RF ML-based classifier(s) by transmitting adversarial perturbations OTA and thus substantially reduce their classification accuracy. To test this capability, adversarial perturbation were broadcast on the X300 SDR, while RadioML signals were transmitted and received on the N200 SDR on separate antennas within LoS of the X300 (this was done on the same SDR because only two SDRs were available). Transmission delays between the two devices were randomized to ensure lack of synchronization. The results are shown in Figure 11. As can be seen, there is only a moderate decrease in efficacy when transmitting the adversarial perturbations on a separate device unsynchronized with the first.

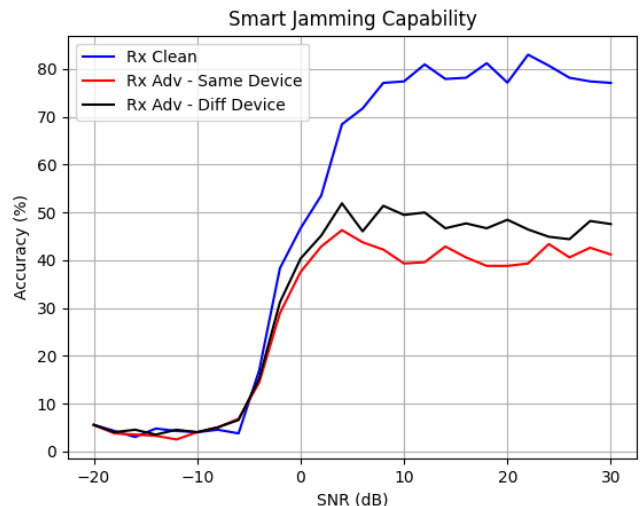

*Figure 11. Smart jamming OTA results comparing classifier accuracy when adversarial perturbations were transmitted from the same device vs different devices.*

## V. DISCUSSION AND CONCLUSION

We considered the use of DL-based algorithms for automatic modulation classification and showed adversarial attacks against such algorithms are both viable and practical. We showed that untargeted CUAAs are significantly more effective than fully universal adversarial attacks, despite adding essentially no additional computational overhead. We did not find a low-latency substitute for sample-specific targeted attacks, and plan to pursue this in future research. Furthermore, to our knowledge, we were the first to broadcast these adversarial attacks over a real LoS channel using SDRs. We found that both the DNN AMC classifier as well as the adversarial attacks were not significantly affected by these real-channel effects. In future work, we will further explore real channel effects by examining NLoS channels and over increasing distances between transmitter and receiver. Finally, we leveraged the time-invariant property of universal and class-specific adversarial attacks to show that these attacks can be broadcast from a separate device than the device transmitting the actual communications signals. We have shown that there is only a minor penalty for this in LoS channel conditions. This capability, which we call smart jamming, can render systems that leverage DNN-based AMC ineffective with much less power than traditional jammers, while not impeding communication between friendly devices.

In most of the work on RF adversarial perturbations, adversarial attack algorithms are adopted from the computer vision literature with minimal modifications. Here, we found it critical to add random phase offsets to the signals when generating the adversarial perturbations for them to be effective on a physical channel. We believe incorporating the nuances of RF communications, including decoding requirements and the nuances of OTA transmission, into the adversarial attack algorithm could further improve results, and we leave this to future work.

In this work, we did not examine defenses against such adversarial attacks. Some previous studies have proposed defenses against these attacks [11,16] and in future work we plan to evaluate the effectiveness of these defenses, as well as others proposed in the computer-vision literature, against our attacks.

### ACKNOWLEDGMENT
The authors would like to thank Joe Storniolo, Sam Friedman, Juan Hodelin, John Chauvin, and Alireza Behbahani for many helpful discussions during the research and preparation of this study.